\documentstyle[12pt]{article}
\newcommand{\koppl}[3]{ \left[ {#1}\otimes{#2}\right]^{[#3]} }
\newcommand{\koppll}[5]
 { \left[ \left[ {#1} \otimes {#2}\right]^{[#3]} \otimes {#4}
                                 \right]^{[#5]} }
\newcommand{\ket}[1]{\left|{#1}\right\rangle}
\newcommand{\braket}[2]{\left\langle{#1}\right.\left|{#2}\right\rangle}
\newcommand{\expect}[1]{\left\langle{#1}\right\rangle}
\newcommand{\bramket}[3]{\left\langle\,{#1}\,\left|\,{#2}\,
            \right|\,{#3}\,\right\rangle}
\newcommand{\brarket}[3]{\left\langle\,{#1}\,\left\|\,{#2}\,
            \right\|\,{#3}\,\right\rangle}
\newcommand{\bqn}{\begin{eqnarray}}
\newcommand{\eqn}{\end{eqnarray}}
\newcommand{\beq}{\begin{equation}}
\newcommand{\eeq}{\end{equation}}
\newcommand{\x}{\times}
\newcommand{\ST}{{\cal S}^T}
\newcommand{\F}{{\cal F}}

\newcommand{\lm}{\lambda}
\newcommand{\k}{\kappa}
\newcommand{\us}{\underline{\sigma}}
\newcommand{\ut}{\underline{\tau}}
\begin{document}
\title{                 Test of Time Reversal Symmetry\\
                                    in the\\
                             Proton Deuteron System}
\author{
                                 Michael Beyer\\
                     Institut f\"ur Theoretische Kernphysik\\
               Universit\"at Bonn, Nussallee 14-16, 5300 Bonn, FRG}
\date{}

\vspace{1.5in}
\maketitle
\vfill
\begin{abstract}
Internal target experiments with high quality proton beams allow for a
new class of experiments providing null tests of time reversal  symmetry
in forward scattering. This could yield more stringent  limits  on
T-odd  P-even observables. A excellent candidate for such experiments is
the proton deuteron system. This system is analyzed in terms of
effective T-violating P-conserving nucleon-nucleon interactions and
bounds on coupling strengths  that  might  be expected are given.

\end{abstract}
\vfill
\newpage

\centerline{\large\bf 1. Introduction}

The existence of CP-violation is well established through decays in the
$K_L$, $K_S$ system [1]. Despite strong efforts, experiments on other
systems  have  given only bounds on CP-violating and on T-violating
interactions.  Both  symmetries are treated equivalently in the
following due to no experimental  evidence  of CPT-violation.

Among these efforts are the measurements of the electric dipole moments
of the neutron [2,3], the atom and the electron [4], T-odd  correlations
in $\beta$-decay [5] and $\gamma$ angular distributions [6],
compound-nucleus  reactions  as  well  as tests of detailed balance
[7-9]. For a review on more  experiments  see  refs. [10-14].

It is important to distinguish P-violating (P-odd) from P-conserving
(P-even, viz. C-odd) breaking of time reversal symmetry. It is only the
P-odd violation of CP that has been established  and  accommodated  by
the  Kobayashi-Maskava matrix. To date there is no experimental evidence
for P-even violation  of  CP symmetry. Also the question whether the
standard  model  alone  provides  any T-odd P-even interaction on tree
level,  is  presently  all  under  discussion [15,16].

Experiments in nuclear systems are usually  analyzed  in  terms  of
effective T-odd nucleon-nucleon (NN)  interactions  [17,18].  This
seems  a  reasonable parameterization for the moderate energies involved
in most  experiments.  For complex nuclei similar to the treatment of
P-violation [19],  effective  T-odd one particle potentials have been
introduced  [20,21].  Due  to  the  complex structure of nuclei,
enhancement factors may occur that lead  to  advantageous experimental
observables [20,22-25]. Many examples of such an enhancement have been
found in the context of parity violation experiments  such  as  in
$^{180}$Hf $\gamma$-correlation or $^{139}$La thermal neutron
transmission, for  a  review  see  e.g. [26]. Therefore, similar
experiments to test time reversal symmetry have  been suggested [24,
27].

Unfortunately, enhancement factors of this kind seem absent for light
nuclear systems, compare also [18]. On the other hand, the proton
deuteron (pd) system considered here would fully utilize the high
luminosity of high quality proton beams  combined  with  internal
target  techniques.  A   forward   scattering experiment would provide a
null test of time reversal symmetry, as pointed out by Conzett recently
[28]. Other setups for scattering processes (two particles in two
particles out) would not lead to a true null  test,  in  the  sense  of
measuring a unique T-odd observable [29]. Forward scattering could lead
to  an experimental accuracy of $|\expect{T-{\rm odd}}| < 10^{-6}
|\expect{T-{\rm even}}|$ [30]. This  accuracy  has been unmatched before
and therefore  this  system  will  be  analyzed  in  the following.

\vspace{1cm}

\centerline{\large\bf 2. T-odd Correlations and Forward Scattering}

The optical theorem relates the total cross section to the forward
scattering amplitude $\F$ via (final state polarization not observed)
[32]
\beq
\sigma_{tot} = 4\pi/k ~{\rm Im}\, tr(\F\rho)/tr(\rho)
\eeq
For convenience the initial state density matrix $\rho$ is  expanded  in
terms  of product spin tensors
\beq
\koppl{S_1^{[\lm]}}{S_2^{[\k]}}{J}_M
= \sum_{\nu\mu} \braket{\lm\mu\k\nu}{JM}S_{1,\mu}^{[\lm]}S_{2,\nu}^{[\k]}
\eeq
with $S_{1,\mu}^{[\lm]}(S_{2,\nu}^{[\k]})$  spin tensor for particle
1(2) [31,  33],  of  rank  $\lm(\k)$  and the Clebsch-Gordan
coefficient     $\braket{\lm\mu\k\nu}{JM}$ [31].  The   spin   tensors
are normalized, in accordance with the Madison  convention  [33],  such
that  the reduced matrix element is given by $\brarket{j}{S^{[\lm]}}{j}
= \hat{j}\hat{\lm}$, where $\hat{j} = (2j+1)^{1/2}$.  The density matrix
then reads ($S^{\dagger\lm}$   denotes the hermitian conjugate)
\beq
\rho = tr(\rho)(\hat{j}_1\hat{j}_2)^{-2}
\sum_{\lm\k}\sum_{JM} {\rm t}_{\lm\k,M}^{[J]}
\koppl{S_1^{\dagger [\lm]}}{S_2^{\dagger [\k]}}{J}_M
\eeq
with ${\rm t}_{\lm\k,M}^{[J]}$     denoting the tensor moments, viz.
\beq
{\rm t}_{\lm\k,M}^{[J]} = \sum_{\nu\mu}\braket{\lm\mu\k\nu}{JM}{\rm
t}_{\lm\mu,\k\nu} = \sum_{\nu\mu} \braket{\lm\mu\k\nu}{JM} {\rm
t}_{1,\mu}^{[\lm]}{\rm t}_{2,\nu}^{[\k]}
\eeq
The last equality in eq. (4) holds for uncorrelated incoming particles.

In the above spin basis eq. (2)  and  in  the  center  of  mass  system
(i.e. ${\bf k} =({\bf p}_1-{\bf p}_2)/2, {\bf p}_1 + {\bf p}_2 =0)$ the
forward scattering amplitude $\F$ may be decomposed  as follows
\beq
\F=\sum_{\lm\k}\sum_{J} \F_{\lm\k;J}
\koppll{S_1^{[\lm]}} {S_2^{[\k]}} {J} {k^{[J]}} {0}
\eeq
with
\beq
k_M^{[J]}=\left[\frac{4\pi J!}{(2J+1)!!}\right]^{1/2} i^L Y_{JM}(\hat{k})
\eeq
where $\hat{k}={\bf k}/k$. Note, that for ${\bf k}=k\hat{e}_z$ (Madison
convention) only moments with  $M=0$, can contribute to the total cross
section  (rotational  symmetry  of  forward scattering). Evaluating the
spin trace, viz.
\beq
tr\left(\rho\
\koppll{S_1^{[\lm]}} {S_2^{[\k]}} {J} {k^{[J]}} {0} \right)
= tr(\rho) \koppl{{\rm t}_{\lm\k}^{[J]}}{k^{[J]}}{0}
\eeq
the total cross section can be written as
\beq
\sigma_{tot}=4\pi/k \sum_{\lm\k}\sum_{J}
{\rm Im} \left(\F_{\lm\k;J}\koppl{{\rm t}_{\lm\k}^{[J]}}{k^{[J]}}{0}\right)
\eeq
This is the most general form of the  total  cross  section,  if  final
state polarization is not observed. If $\k=0 (\lm=0)$ and therefore
$\lm=J (\k=J)$  in eq. (8), then particle 2(1)  is  not  polarized
and  for  $J\neq 0$  the  forward scattering amplitude $\F_{J0;J}$
($\F_{0J;J}$) depends on the polarization of one particle only. If
$\k\neq 0$ and $\lm\neq 0$, then  the  total  cross  section  depends
upon  the correlation between the spins of the two incoming particles.

To exhibit the symmetry relations of $\F_{\lm\k;J}$, the forward
scattering  amplitude will now be decomposed according to parity and
time  reversal  symmetry.  This follows from the properties of the spin
tensors and  the  spherical  harmonics under parity and time reversal
symmetry [31], viz.
\beq
\F_{\lm\k;J} = \frac{1}{4}\sum_{\eta\tau}(1+\eta\tau(-1)^{\lm+\k})
             (1+\tau(-1)^{\lm+\k+J}) F_{\lm\k;J}^{\eta\tau}
\eeq
with $\eta = 1(-1)$ for P-even(odd)  and $\tau=1(-1)$  for  T-even(odd)
symmetry.  For $\lm=\k=0$, i.e. unpolarized initial particles, only
$\eta=\tau=1$ can contribute  to  the total cross section. Inserting
$\lm = 0$, $\k=J$ or $\k=0$, $\lm=J$ and $\tau=\pm 1$ in  eq. (9),
one finds that in forward scattering only  T-even  ($\tau=1$)
amplitudes  are possible. This situation is different from  pure
P-violation  ($\eta=-1$, $\tau=1$  in eq.(9)), which means that a P-odd
experiment can be  conducted  by  polarizing only one of the
participating particles (projectile or target).  To  test  for time
reversal symmetry with a  single  measurement  one  needs  to  have both
particles polarized and measure a  correlation  of  the  spins. This
can  be achieved by setting $\k\neq 0$ and $\lm\neq 0$ in eq.(9). Then
there is still the  choice  to distinguish between P-even (J  even)  or
P-odd (J  odd)  violation  of  time reversal symmetry. This way all
possible observables  for  which  final  state polarizations are not
observed are exhausted. In the following a notation  for
$\F_{\lm\k;J}^{\eta\tau}$ that exhibits the symmetry relations more
obviously is used, viz. $\F_{\lm\k;J}^{E}$, $\F_{\lm\k;J}^{P}$,
$\F_{\lm\k;J}^{T}$, $\F_{\lm\k;J}^{TP}$ for the T-even P-even (E), T-even
P-odd (P), T-odd P-even (T), T-odd P-odd (TP) forward scattering
amplitude. See also table 1.

\begin{table}[t]
\caption{
Symmetry relations implied on the forward  scattering
amplitude  via eq.(9)}
\[
\begin{array}{c|cc|}
&J~{\rm even}&J~{\rm odd}\\
\hline
(\lm+\k)~{\rm even}&E&TP\\
(\lm+\k)~{\rm odd} &T&P
\end{array}
\]
\end{table}

For spin-$\frac{1}{2}$ particles ($j_1 =j_2 =\frac{1}{2}$), the
following amplitudes contribute to forward scattering: $\F_{00;0}^E$,
$\F_{10;1}^P$, $\F_{01;1}^P$, $\F_{11;0}^E$, $\F_{11;2}^E$,
$\F_{11;1}^{TP}$. Note that  there  is  no T-odd P-even forward
scattering amplitude $\F$. For the more general  case  with final momentum
{\bf k}'$\neq${\bf k}, all possible  spin  combinations  have  been
given  by Wolfenstein [34].

For the pd-system with  $j_1 =\frac{1}{2}$ and  $j_2 =1$,  the
amplitude  has  a  much  richer structure. Beside the amplitudes
mentioned above additional spin contributions are possible:
$\F_{02;2}^E$, $\F_{12;1}^P$, $\F_{12;3}^P$, $\F_{12;2}^T$. For the more
general case {\bf k}$\neq${\bf k}' the amplitudes have been given by
Seyler [35]. The  symmetry  assignments  can  be easily reproduced with
help of table 1. For this system in particular a  T-odd P-even amplitude
is present, which would be zero, if time reversal  invariance holds. An
experiment sensitive to $\F_{12;2}^T$ would be a  true  null  test  of
time reversal invariance.

To be more specific, the  tensor  moments  ${\rm t}_{\lm\k}^{[J]}$,  eq.
(8), of  the  incoming particle may be chosen such that the total cross
section for  spin-$1/2$  spin-1 scattering is given through ({\bf
k}=$k\hat{e}_z$)
\beq
\sigma_{tot}^T = \sigma_{tot}^0 -\frac{4\pi}{k}
\sqrt{\frac{2}{15}}
{}~{\rm Im}\left(\F_{12;2}^T {\rm t}_{12}^{[2]}\right)
\eeq
with $\sigma_0$ the total unpolarized cross section. The tensor moment
${\rm t}_{12}^{[2]}$  may  be rewritten in terms of cartesian
coordinates, by using eq. (4) for uncorrelated inital states and
rewriting ${\rm t}_{\pm 1}^{[1]} = \mp(P_x\pm iP_y)/\sqrt{2}$  and ${\rm
t}_{\pm 1}^{[1]} = \mp(P_{xz}\pm iP_{yz})/\sqrt{3}$ Here, $P_y$  and
$P_x$  denote the polarizations and $P_{xz}$, $P_{yz}$  the  alignment
of  the initial particles with respect to the cartesian basis, viz.
\beq
{\rm t}_{12}^{[2]} = -i(P_{xz}P_y - P_{yz}P_x)/\sqrt{3}
\eeq
Note, that due to rotational symmetry  of  forward  scattering  the
cartesian coordinate system may  be  rotated  around  the  z-axis,  such
that $P_{yz}P_x =0$. Inserting eq. (11)  into  eq.  (10),  a  T-odd
P-even  "total  cross  section asymmetry" or "total spin correlation"
$\ST$  may then be defined through
\beq
\sigma_{tot}^T =\sigma_{tot}^0 (1+P_{xz}P_y\ST)
\eeq
which is explicitly given by
\beq
\ST = \frac{4\pi}{3k} \sqrt{\frac{2}{5}}~ {\rm Re}(\F^T_{12;2})/\sigma_{tot}^0
\eeq
Note that being T-odd, the correlation $\ST$  is sensitive to the {\em
real} part of the forward scattering amplitude $\F^T_{12;2}$, since
${\rm t}^{[2]}_{12}$ is  imaginary,  eq.  (11).  Eq. (13) will be
evaluated in the following for the proton-deuteron system.  Note, that
the tensor structure
$\koppll{S_1^{[1]}} {S_2^{[2]}} {2} {k^{[2]}} {0}$
implied here is present  in the alternative expression
$(\underline{\sigma}\x {\bf J}\cdot {\bf k}) ({\bf J}\cdot \hat{k})$
with $\us={\bf S}_1$  and ${\bf J} = {\bf S}_2$,
sometimes used in this context.

The dependence on the polarization of the incoming particles eqs.
(10),(12) is unique for a T-odd P-even observable. The asymmetry may be
extracted from  the total cross section in a transmission experiment by
switching the sign of  $P_y(P_{xz})$ while keeping $P_{xz}(P_y)$
constant and subtract  the  transmission  factors corresponding to the
change of sign [28,30].

\vspace{1cm}
\centerline{\large\bf
                         3. Nucleon Nucleon Amplitudes}

Now the important question arises, which types of two nucleon amplitudes
$t_{NN}$ will lead to a T-odd P-even observable in the pd-system. In
general also three body forces could contribute. However, though they
may  be  present  they  are presumably not the dominant forces and are
neglected in the following.

Due to the energy regime considered here, the question  raised  above
may  be answered  in  lowest  order  rescattering  series.   Indeed,
comparison   of experiments show that the total pd cross section is
roughly the sum of neutron proton and proton-proton total cross sections
[36], with  sufficient  accuracy for the present investigation.

As an example the basic mechanisms will be demonstrated  on  the  pd
break-up cross section in some detail. For simplicity I use channel
notation  for  the three nucleon system, i.e. $t_k := t_{NN}(ij)$ with
ijk  permutation  of  particles 123, and following ref. [37]  the
break-up  cross  section  in  lowest  order rescattering approximation
may be written as
\beq
\sigma_{b-up} = 4\frac{E_pE_d}{Ek} tr\left(\frac{\rho}{6} \int
\bramket{\phi_1 {\bf k}_1}{t_2^\dagger + t_3^\dagger}{\phi_0^S}
\bramket{\phi_0^S}{t_2 + t_3}{\phi_1 {\bf k}_1} \delta(E-E_0)\right)
\eeq
with $E = E_p  + E_d$  the total scattering energy, $E_p , E_d$  the
proton  and  deuteron energies resp., $E_0$  the energy of the free
three  particle  state $\phi_0^S$ that  is properly  symmetrized  in
one  pair  coordinate.  Since  all  particles   are identical, the
initial state $\ket{\phi_1{\bf k}_1}$, with the deuteron wave  function
$\phi$,  has been chosen to be in channel 1 [37]. The integral runs over
all continuous and discrete quantum numbers of the final state
$\phi_0^S$. The factor 1/6  takes  account of the 6 fold overcounting
due to the symmetry of the final state. Note,  that due to the rank of
the spin tensor $S_2^{[2]}$ appearing  in  the  T-odd  observable $\ST$
only channels with different indices, viz. $\propto\expect{t_2^\dagger
t_3}$, are nonzero. The direct channels viz. $\propto\expect{t_2^\dagger
t_2}$, $\expect{t_3^\dagger t_3}$, are excluded by spin selection rules.

For $t_{NN}   = t_{NN}^E + t_{NN}^T + t_{NN}^{TP} + t_{NN}^P$ eq. (14)
separates into a  sum  of  terms  with different symmetry properties
under T and P. Then the T-odd P-even total  spin correlation $\ST$ ,
eq. (13) may arise through the following combinations
\bqn
\ST &\propto& \expect{t^T_{NN}t^E_{NN}}/\sigma^0_{tot}\\
\ST &\propto &\expect{t^{TP}_{NN}t^P_{NN}}/\sigma^0_{tot}
\eqn
Note that in the three body system both types, viz. P-even $t^P_{NN}$
and P-odd $t^{TP}_{NN}$, violation of time reversal symmetry may
contribute to  a  measurement of  $\ST$ . They cannot be
distinguished.

However, the bounds on $t_{NN}^{TP}$ are rather stringent from  electric
dipole  moment measurements [2,3,15,38], viz. $|\expect{t_{NN}^{TP}}|<
10^{-10}...10^{-11}|\expect{t_{NN}^{E}}|$.  Also,  the parity violating
amplitude $t_{NN}^P$   is expected be in the  range  of  typical  weak
amplitudes, viz $|\expect{t_{NN}^P}|\simeq 10^{-7}|\expect{t_{NN}^E}|$
[29].  Therefore,  the  combination  of such two body amplitudes would
very likely lead  to  bounds  of $|\expect{t_{NN}^{TP}t_{NN}^P}|\leq
10^{-17}\sigma^0_{tot}$ for the proton  deuteron system.  This  value
is  far  below  the resolution that might be reached in a measurement of
$\ST$.

On the other hand, experimental bounds on  the  strength  of  the
alternative combination  $\expect{t_{NN}^Tt_{NN}^E}/\sigma^0_{tot}$,  eq.
(15), which  provides  a  test  of  generic T-violating P-conserving two
body interactions,  are  much  weaker.  However, comparison of different
experiments is rather difficult, which is  mostly  due to our lacking
knowledge of a  P-even  breaking  mechanism  of  time  reversal
symmetry. This is different from  P-odd  breaking,  where  mechanisms
can  be parameterized in terms of the standard model. In this sense each
experiment is unique, and different experiments can only be  compared by
using  even  mild model assumptions to treat the dynamic behavior of the
system  in  question. Before preceding, the bounds implied by these
experiments will be discussed in the following.

Experiments on complex nuclei are  usually  analyzed  in  terms  of
effective one-body potentials with strength $G^T$ , as suggested by
Michel in the context of parity violation [19]. The probably most
stringent limit  from  $\gamma$-decay  comes from   the   experiment on
$^{57}$Fe   [6].   It   gives   a   bound    through $G^T\simeq (0.7\pm
1.6\pm 0.5)\x 10^{-5} $[20],  where  the  first   error   relates to the
experimental error and the second to different residual interactions in
the shell model analysis. Detailed balance experiments give bounds on
T-odd P-even observables based on statistical interpretation of the
level  distribution  of the  compound  nucleus. In a  most  recent
analysis  Harvey  et  al.   give $G^T <  2.6\x 10^{-4}$ [9].
However, bounds in terms of $G^T$  still include  some nuclear physics
aspects which may change, through not dramatic, bounds given in  terms
of two body amplitudes.

Alternatively, bounds on effective T-odd P-even NN interactions may be
related to observables from T-odd P-odd experiments, assuming that the
P-odd part is due to the standard model. A rough estimate of the
corresponding limit on the T-odd P-even part $f^T$ in the electric
dipole moment of the neutron would be $|f^T|< 2\x 10^{-5}$
[14,38].  From this number implications on the limits on a generic
T-odd P-even meson nucleon coupling strength $g_{MNN}^T$ may be found.
An educated "guess" gives $|g_{MNN}^T/g_{MNN}^E|=\phi<10^{-4}$ [14].
Using essentially the same experimental input Khriplovich argues that
one might "expect" a bound of $|\expect{t_{NN}^T}| < 2\x
10^{-6}|\expect{t_{NN}^E}|$ [38]. The large discrepancy between these
two numbers is mostly due the implicit mass scale, introduced to
express the bound. Khriplovich relates the coupling strength to the
small pion mass, whereas Herczeg's result accommodates for a large
mass responsible for a short range effective T-odd P-even NN
interaction [14]. Therefore, in order to compare the number given in
[38] with the results for the pd scattering given below, it is
necessary to renormalize the mass scale and introduce an additional
factor $(m_A^2/m_\pi^2\simeq 90)$ with $m_A$ the mass of the axial
vector meson given below. One then finds $|\expect{t_{NN}^T}|<2\x
10^{-4}|\expect{t_{NN}^E}|$, in accordance with ref. [14].

A general parity and time reversal symmetry conserving as well  as
rotational and isospin invariant amplitude $t^E_{NN}$ for two nucleons
may be written  in  terms of a Wolfenstein type parameterization [34],
i.e. in the cm frame
\bqn
t^E_{NN} &=& a + b  \us_i\cdot\us_j
               + c~ i(\us_i+\us_j)\cdot{\bf q}\x{\bf p}/m_p^2\nonumber\\
&&             + e  (\us_i\cdot{\bf p}\us_j\cdot{\bf q}
                     -\us_i\cdot\us_j{\bf q}^2/3)/m_p^2\nonumber\\
             &&+ d  (\us_i\cdot{\bf p}\us_j\cdot{\bf q}
                     -\us_i\cdot\us_j{\bf q}^2/3)/m_p^2
\eqn
with momenta ${\bf q} = {\bf k}'-{\bf k}$, ${\bf p} = ({\bf k}'+{\bf
k})/2$ and $m_p$   the  proton  mass introduced  for dimensional
reasons. The functions $a$ to $e$ are functions of Lorentzscalars,  in
the cm system given by ${\bf k}^2$, ${{\bf k}'}^2$  or ${\bf k}\cdot
{\bf k}'$. In the following the first  two  terms are denoted as
scalars, the third term as spin-orbit term, and  the  last  two terms as
tensor terms. Note, that for ${\bf k}={\bf k}'$ the  amplitudes  $a$,
$b$  and  $e$  are respectively related to the amplitudes $\F_{00;0}^E,
\F_{11;0}^E, \F_{11;2}^E$, defined above.

A generic time reversal violating  but  parity  conserving amplitude
$t^T_{NN}$ may be written as follows (see also ref. [37])
\bqn
t^T_{NN}&=& f \us_i\cdot\us_j{\bf q}\cdot{\bf p}/m_p^2
           +g \us_i\x\us_j\cdot{\bf q}\x {\bf p}/m_p^2\nonumber\\
&&         +h ( \us_i\cdot{\bf p}\us_j\cdot{\bf q}
           +   +\us_j\cdot{\bf p}\us_i\cdot{\bf q}
               -2\us_i\cdot\us_j{\bf p}\cdot{\bf q})/m^2_p\nonumber\\
&&         +g'(\us_i-\us_j)\cdot{\bf q}\x{\bf p}/m_p^2 (\ut_i\x\ut_j)_0
\eqn
The functions $f,g,g'$ and $h$ are again functions of  Lorentzscalars.
The  first term is of scalar type, the second depends on spins and
angular momentum,  the third is of tensor type (with arbitrary
dependence on isospin), and  the  last term is of isovector spin-orbit
type,  included  to  account  for  a  possible violation of C-symmetry.
Note, that the T-odd  P-even amplitude  vanishes  for ${\bf k}'={\bf
k}$, which reflects the results for $j_1=j_2=\frac{1}{2}$given above.

Possible NN potentials for the  T-odd  amplitude  of  eq.  (18)  are
provided through ($\rho$-meson) vector exchange, with a C-violating (and
hence  T-violating) isospin dependence [18] or through axial vector
exchange  [18,39],  reflecting the different behavior of axial vector
and pseudo  tensor  interactions  under time reversal symmetry,
respectively. They are given in the following.

The vector exchange potential reads in momentum space, up to order
$p^2/m^2_p$  [18]:
\beq
v_\rho^T=i\phi_\rho\k_\rho g^2_{\rho NN}({\bf q}^2)/
          [8m^2_p({\bf q}^2+m_\rho^2)]
          (\us_i-\us_j)\cdot{\bf q}\x{\bf p}
          (\ut_i\x\ut_j)_0
\eeq
The strength $\phi_\rho=g^T_{\rho NN}/g_{\rho NN}$ denotes the strength
of the T-violating  potential relative to the T-conserving one. The
other parameters  are  taken  from  ref. [40] Table 5, viz.
$m_\rho=0.769$GeV, $\k_\rho=f_{\rho NN}/g_{\rho NN}=6.1$, $g^2_{\rho
NN}({\bf q}^2=m^2_\rho)/4\pi=0.81$, and the ${\bf q}^2$-dependence of the
strength is given by the form factor
\beq
F_{\mu NN}({\bf q}^2) = [(\Lambda^2_\mu - m^2_\mu)/
                         (\Lambda^2_\mu - {\bf q}^2)]^{n_\mu}
\eeq
with $\Lambda_\rho=2$GeV and $n_\rho=2$.

The axial vector exchange potential
reads  in  momentum  space,  up  to  order
$p^2/m_p^2$:
\beq
v_A^T=i\phi_A g^2_{ANN}({\bf q}^2)/
          [8m^2_p({\bf q}^2+m_A^2)]
          ( \us_i\cdot{\bf p}\us_j\cdot{\bf q}
           +\us_j\cdot{\bf p}\us_i\cdot{\bf q}
           -\us_i\cdot\us_j{\bf q}\cdot{\bf p})
\eeq
with $m_A=1.26$GeV axial vector  meson  mass.  The  isospin  dependence
is  not restricted and may be isoscalar, isovector or isotensor. The
coupling strength $g_{ANN}$ is  not  well  know  empirically  from  NN
potentials.  One  may  choose $g^2_{ANN}({\bf q}^2=m_A^2)/4\pi=3.8$ [39]
for an isoscalar $\ut_i\cdot\ut_j$-dependence,  with  a  cut-off similar
to the one used in the Bonn potentials with $n_A=1$ and $\Lambda=2$GeV
as for other heavy mesons [40]. The coupling constant would then be
$g_{ANN}^2({\bf q}^2=0)/4\pi=1.56$, close to the values of the
$a_1$-meson of the Virginia potential with meson nucleon coupling
constants calculated from  quark  symmetry  rules  and  quark models
[41].

\vspace{2cm}
\centerline{\large\bf
                                  4. Results}

The deuteron wave function used in the calculation is generated  by  the
Bonn potential [42]. To simplify the  calculation  of  spin  matrix
elements  only S-waves are considered.

The potential to determine the functions $a$ to $e$ for the strong
amplitudes  in eq. (17) are provided through one boson exchange with the
parameters  of  the Bonn potential [40]. However,  for  the  present
estimate  the  amplitude  is evaluated in Born approximation  only,
viz.  $t^E_{NN}\simeq v_E$ in  eq.  (16).  This approximation  may  not
be  sufficient   but   simplifies   the   calculation considerably.
However, any error in  the  strong  transition  amplitude  would effect
the T-odd observable only linearly, see eq. (15). A comparison with the
T-even break-up data at the relevant energies shows, that a  factor  of
three uncertainty for the transition amplitude, and hence the T-odd
observable,  is possible. A more elaborate analysis is certainly needed
to reduce  the  number of possible uncertainties. Some uncertainties
however  may  also  be  due  to relativistic effects, which should not
be negligible above momenta larger than 1 GeV. Part of that is taken
into account  by  using  relativistic  kinematics instead of the
nonrelativistic one.

Presently, only elastic and break-up  channels  are  considered.  Other
final channels including particle production (e.g. of pions) are
neglected. This  is justified as long as the energies are  below
particle  threshold  (e.g.  pion threshold). However  open  channels
may  become  more  important  for  higher energies. Fortunately, any
additional contribution is in favor of  the  P-even T-odd observable
considered, if  no  accidental  cancellations  occur  at  the energy
chosen for the experiment.

The first model (model I) considered is a estimate of the bounds that
might be expected from a pd-forward scattering experiment.  To  this
end  the  unknown functions $f, g, h$ and $g'$ in eq. (18) are taken
constant.  For  a  very  short range interaction and moderate energies
this  may  be  justified,  since  the deuteron wave function  cuts  out
contributions  of  higher  momenta  in  the integrals and therefore the
momentum dependence is not so important.

For the elastic channel the isospin independent amplitudes dominate.
However, the contribution of this channel to the total spin correlation
coefficient  is negligible, i.e. two orders of magnitude smaller than
the one induced  by  the break-up channel, which will be discussed in
the next  paragraph.  The  charge dependent  amplitude  would
contribute  only  via  charge  symmetry  breaking transitions and
therefore would lead to a suppression of the amplitude by  the size of
charge symmetry breaking, i.e. $\simeq 10^{-3}$ see e.g. [43].

Contributions due to the break-up channel of model I  are  shown  in
fig.  1. Since the total phase is unknown only absolute values are
shown.  The  largest contribution is a tensor T-odd amplitude ($\propto
h$ in eq.(18)) with a scalar T-even rescattering  ($\propto b$  in  eq.
(17)),  generated  by  $\pi$, $\rho$ exchange.   Other contributions in
this simple estimate  are  roughly  one  order  of  magnitude smaller.

To obtain the total spin correlation $\ST$, one needs  to  know  the
unpolarized {\em total} cross section $\sigma_{tot}^0$. Unfortunately
not all  relevant  energy  values  are deducible from experimental data
[36]. There is a gap between $k_{cm}\simeq 0.2$GeV and $k_{cm}\simeq
0.5$GeV. In fact, as seen in both figures this is the  region  where
the maxima of the break-up part of $\ST\sigma_{tot}^0$ occur. Provided a
measurement of $\sigma^0_{tot}$ in that region leads to
$\sigma^0_{tot}\simeq 50$mb, which is the value at $k_{cm}\simeq
0.5$GeV [36], and provided an accuracy of $10^{-6}$ as mentioned in the
introduction, a  rough  bound on $\expect{t^T_{NN}}$ could be
achieved from model I to be
$|\expect{t^T_{NN}}|<2\x10^{-4}|\expect{t^E_{NN}}|$.

The second approach (model II) leads to bounds  on  coupling  strength
rather than to bounds on matrix elements. This is one step further in
the  analysis. To this end, the unknown functions $f, g, g'$ and $h$ of
eq. (18) need to be related to the T-violating potentials given in eq.
(19) and (21). Due to the  weakness of the T-odd interaction the two
body amplitude  will  be  evaluated  in  Born approximation, i.e.
$t^T\simeq v^T$  in eq. (16). Therefore, the T-violating potentials does
not give terms proportional to $g$.

In this framework  the  dominant  contribution  relevant  for  elastic
proton deuteron scattering comes from the tensor part of an axial
vector ($f_1$ -meson, eq.(21)) T-violating interaction interfering with
the spin  orbit  part  of  a ($\omega$-, eff. $\sigma$-meson)
T-conserving rescattering  interaction.  Isospin  dependent meson
exchange contributes only via charge symmetry breaking. Nevertheless, as
in the more general case, the elastic contribution can be  neglected
compared to the break-up contribution.

For the break-up channel the dominant contribution  in  model  II  is
through charged meson exchange potentials. This is due to the spin
orbit  part  of  a T-violating vector exchange ($\rho^\pm$ meson, eq.
(19)) with a tensor ($\pi^\pm$-, $\rho\pm$-meson) T-conserving
rescattering interaction. The result is shown as long dashed line in
fig. 2. Since the T-odd coupling strength $\phi_\rho$  is not known
$\ST\sigma^0_{tot}$ is  given in units of $\phi_\rho$. Other lines
represent contributions  due  to  the  T-violating axial vector exchange
with a T-conserving rescattering contribution induced by $\pi$, $\rho$
exchange. The solid, dashed and dotted line are due to  tensor
(scalar), scalar (tensor) and tensor (tensor)  T-violating
(T-conserving  rescattering) interaction. The total spin correlation
$\ST$  is given in units  of  the  unknown strength $\phi_{a_1}$.

The suppression of the solid and dashed lines in fig. 2 (model II)
compared to fig. 1 (model I) is partly due to a  factor  1/3
(Clebsch-Gordan-coefficient) arising from the evaluation of the scalar
contribution. Note  also,  that  the results of model II drop faster
with momentum transfer than those of model  I. This is, since short
range correlations are treated more properly in model  II due to the
form factor cut off. In addition, not all possible contributions of
model I are covered by the more microscopic approach of model  II,
since  the meson exchanges are considered in Born approximation only.

The maxima of the T-odd asymmetry in fig. 2 appear at $k_{cm}\simeq
0.3$GeV.  They  are $\ST\sigma^0_{tot}\simeq 0.04\phi_\rho$mb for
$\rho$-exchange, and $\ST\sigma^0_{tot}\simeq 0.02\phi_{a_1}$mb  for
$a_1$ exchange. Provided the experimental accuracy of $10^{-6}$ and
$\sigma^0_{tot}\simeq 50$mb, a bound on $\phi_\rho$ may be achieved of
$\phi_\rho=g^T_{\rho NN}/g_{\rho NN}<10^{-3}$   and a bound on
$\phi_{a_1}$  of $\phi_{a_1} =  g^T_{ANN}/g_{ANN}< 2\x 10^{-3}$. Note,
that these bounds are on T-odd coupling  constants  and  not  on matrix
elements $\expect{t^T_{NN}}$.

\vspace{1cm}
\centerline{\large\bf
                         5. Discussion and Conclusion}

Firstly,  it  is  important  to   note   that   microscopic   models
for   a parameterization of T-odd P-even forces lead to nonzero
contributions  to  pd forward scattering amplitude. This result  is,
presumably,  not  affected  by final state interaction effects.  The
reason  is  that  it  is  not  a  phase measurement but measures the
total  T-odd  cross  section.  Indeed,  using  a different terminology,
final state interaction is responsible to  exhibit  the generic T-odd
effect.

Also, Coulomb interaction does not lead to  divergences  for  the  T-odd
spin correlation. This is due to the necessary interference of  the
Coulomb  force with a T-odd interaction. Therefore, Coulomb-interaction
occurs only  linearly in the observable.  Due  to  the  weakness  of
the  electromagnetic  coupling constant, Coulomb interaction can be
neglected in the present treatment.

In this paper I present a first estimate on the bounds of  T-violating
P-even potentials that might be expected from a pd forward scattering
experiment.  It seems that the favored cm momentum to conduct an
experiment is  around $k_{cm}\simeq 0.3$GeV/c ($k_{lab}\simeq
0.5$GeV/c). However, not all possible  T-odd  channels  have been taken
into  account  so  far,  i.e.  the  pion  production  channels  are
neglected. Conclusions might change qualitatively,  if  those  channels
would dominate the T-odd observable; in particular higher energies might
become more preferable. In order to reduce  possible  uncertainties  a
next  step  should incorporate the  complete  two  body  t-matrix  of
strong  interaction.  Some relativistic effects have  been  taken  into
account  by  using  relativistic kinematics. However, lacking a
covariant theory of three interacting particles the analysis at even
higher energies may become more difficult.  In  addition, the total
cross section needs to be known experimentally in the energy  region
where the T-odd observable has its maximum. It may  also  be
advantageous  to investigate  other  spin-1  nuclei  than  the
deuteron,  which  may  lead  to enhancement through collective effects,
even  if  the  polarization  of  such nuclei is more delicate.

Although the theoretical  analysis  can  still  be  improved,  it  is
already possible to conclude that carrying out a forward scattering
experiment to test time  reversal  symmetry  is  highly  desirable.  It
will  provide  a  direct measurement of bounds on T-odd  P-even NN
amplitudes  of $|\expect{t^T_{NN}}| < 2\x 10^{-4} |\expect{t^E_{NN}}|$
with an accuracy compatible to the  measurements  of  electric  dipole
moments. It has been shown that due to the relatively simple system the
bounds that may be reached on that fundamental symmetry can directly  be
related  to bounds on coupling constants of  generic  effective
T-violating  P-conserving model NN interactions and would lead to $\phi
< 10^{-3}$.

I gratefully acknowledge very  useful  discussions  with  P.D.
Eversheim,  F. Hinterberger, W. Gl\"ockle and B.C. Metsch.

\newpage
\centerline{\large\bf Figure Caption}
\vspace{1cm}
\centerline{\bf Figure 1}
Model I: Total spin correlation $|\ST\sigma^0_{tot}|$ due to the
break-up  channel: solid line contributions due to $\expect{hb}$, dashed
line due  to $\expect{fd}$,  long  dashed line due to $\expect{g'd}$,
dashed dotted line due to $\expect{gc}$, dotted line  due  to
$\expect{hc}$. See eqs. (16,17,18).

\centerline{\bf Figure 2}
 Model II: Total spin correlation $|\ST\sigma^0_{tot}|$  due to the
break-up  channel: long dashed line $\rho$ exchange, all other lines due
to  $a_1$   exchange  T-violating P-conserving potential. Solid line
scalar(tensor) dashed line  tensor(scalar), dotted   line
tensor(tensor)    T-violating    interaction    (T-conserving
rescattering). Results are normalized to the T-odd strength parameter
$\phi_\rho, \phi_{a_1}$, resp.

\newpage
\centerline{\large\bf
References           }
\begin{enumerate}
\itemsep0ex
\item J.H. Christenson, J.W. Cronin, V.L. Fitch and R. Turlay, Phys.
      Rev. Lett. 13 (1964) 138
\item I.S. Altarev et al., Phys. Lett. B102 (1981) 13
\item N.F. Ramsey, Phys. Rep. 43C (1977) 409;\\
      N.F. Ramsey, Ann. Rev. Nucl. Sci. 32 (1982) 211
\item S.K. Lamoreaux, Nucl. Instrum. Methods Phys. Res. Sect. A 284
      (1989) 43
\item M.B. Schneider,  F.P.  Calaprice,  A.L.  Hallin,  D.W.  Mac Athur,
      D.F. Schreiber, Phys. Rev. Lett. 51 (1983) 1239\\
      A.L. Hallin, F.P.  Calaprice,  D.W.  MacAthur,  L.E.   Piilonen,
      M.B. Schneider, D.F. Schreiber, Phys. Rev. Lett. 52 (1984) 337
\item N.K. Cheung, H.E. Henrikson and F. Boehm, Phys. Rev. C16 (1977)
      2381  and references therein\\
      J.L. Gimlett, H.E. Hendrikson, N.K. Cheung and F. Boehm, Phys.
      Rev. Lett. 42 (1979) 354
\item J.B. French, V.K.B. Kota, A. Pnadey and S. Tomsovic, Phys. Rev.
      Lett.  58 (1987) 2400\\
      J.B. French, A. Pandey and J. Smith, in Tests of time reversal
      invariance in  neutron  physics,  eds.  N.R.  Roberson  et  al.
      (World   Scientific Publishing, Singapore, 1987) p. 80
\item D. Boos\'e, H.L. Harney and H.A. Weidenm\"uller, Phys. Rev. Lett.  56
      (1986) 2012\\
      V.E. Bunakov, E.D. Davis, H.A. Weidenm\"uller, Phys. Rev. C 42
      (1990) 1718
\item E. Blanke, H. Driller, W. Gl\"ockle, H. Grenz, A.  Richter,  G.
      Schrieder, Phys. Rev. Lett. 51 (1983) 355\\
      H.L. Harney, A. Hpper, A. Richter, Nucl. Phys. A518 (1890) 35
\item L. Wolfenstein, Ann. Rev. Nucl. Part. Sci. 36 (1986) and ref.
      therein
\item E.M. Henley, Montreal 1989, Proceedings International Symposium on
      Weak and Electromagnetic Interaction in Nuclei,  ed.  P. Depommier
      (Editions Frontieres, Gif-sur-Yvette, 1989) p. 181
\item K. Kleinknecht, Ann. Rev. Nucl. Sci. 26 (1976) 1
\item R.G. Sachs, The Physics of Time Reversal (University  of  Chicago
      Press, Chicago and London, 1987)
\item P. Herczeg, in Tests of time reversal invariance in neutron
      physics, eds. N.R. Roberson et al. (World Scientific Publishing,
      Singapore, 1987) p. 24
\item P. Herczeg, Hyp. Int. 43 (1988) 77
\item P. Herczeg, private communication\\
      M. Simonius, private communication
\item P. Herczeg, Nucl. Phys. 75 (1966) 655
\item M. Simonius, Phys. Lett B58 (1975) 147   \\
      M. Simonius and D. Wyler Nucl. Phys. A286 (1977) 182
\item F.C. Michel, Phys. Rev. 133B (1964) 3329
\item M. Beyer, Nucl. Phys. A493 (1989) 335
\item R.J. Blin-Stoyle and F.A. Bezerra-Coutinho, Nucl. Phys. A211
      (1973) 157
\item W.C. Haxton and E.M. Henley, Phys. Rev. Lett. 51 (1983) 1937
\item A. Griffith and P. Vogel, Phys. Rev. C43 (1991) 2844
\item V.E. Bunakov, Phys. Rev. Lett. 60 (1988) 2250
\item L. Stodolsky, Phys. Lett. 172B (1986) 5
\item E.G. Adelsberger and W.C. Haxton, Ann. Rev. Nucl. Part. Sci. 35
      (1985) 501
\item P.K. Kabir, Phys. Rev. Lett. 60 (1988) 686 \\
      P.K. Kabir, Phys. Rev. D37 (1988) 1856
\item H.E. Conzett, LBL Berkley 1992, LBL-31929\\
      H.E. Conzett, private communication
\item F. Arash, M.J. Moravcsik, G.R. Goldstein, Phys. Rev. Lett. 54
      (1985) 2649
\item P.D. Eversheim et al., Phys. Lett. B 256 (1991) 11, and private
      communication
\item A.R. Edmonds, Angular Momentum in Quantum Mechanics (Princeton
      University Press, Princeton  1974)
\item R.J.N. Phillips, Nucl. Phys. 43 (1963) 413
\item M. Simonius, in  Procedings  of  the  third  international
      symposium  on polarization phenomena in  nuclear  reactions
      (University  of  Wisconsin Press, Madison, 1971) p. 401\\
      M. Simonius, in Lecture Notes in Physics vol.  30:  Polarization
      Nuclear Physics (Springer-Verlag, Berlin Heidelberg, 1974) p. 38
\item L. Wolfenstein, J. Ashkin, Phys. Rev. 85 (1952) 947
\item R.G. Seyler Nucl. Phys. A124 (1969) 253
\item Landolt-B\"ornstein, New Series I/12b
\item W. Gl\"ockle, The Quantum Mechanical  Few-Body  Problem
      (Springer-Verlag, Berlin Heidelberg, 1983)
\item I.B. Khriplovich, Nucl. Phys. B352 (1991) 385
\item E.C.G. Sudarshan, Proc. Roy. Soc. A305, (1968) 319
\item R. Machleidt, K. Holinde and Ch. Elster, Phys. Rep. 149 (1987) 1
\item M. Bozoian and H.J. Weber, Phys. Rev. C28 (1983) 811\\
      M. Beyer and H.J. Weber, Phys. Rev. C 35 (1987) 14
\item M. Fuchs, Bonn, private communication
\item M. Beyer and A.G. Williams, Phys. Rev. C 38 (1988) 779 and ref.
      therein
\end{enumerate}

\end{document}